\documentclass[prd,twocolumn,superscriptaddress,floatfix,amsmath,amssymb]{revtex4-1}
\usepackage[english]{babel}
\usepackage{graphicx}
\usepackage{dcolumn}
\usepackage{bm}
\usepackage{verbatim}
\usepackage{mathrsfs}
\usepackage{cancel}
\usepackage{epstopdf}
\usepackage{color}

\newcommand{\ii}{\mathrm{i}}

\usepackage[usenames,dvipsnames]{pstricks}
\usepackage{epsfig}
\usepackage{pst-grad} 
\usepackage{pst-plot} 

\usepackage{hyperref}

\begin{document}

\title{Universality and thermalization in the Unruh Effect}
\author{W. G. Brenna}
\email{wbrenna@uwaterloo.ca}
\affiliation{Department of Physics and Astronomy, University of Waterloo, Waterloo, Ontario N2L 3G1, Canada}
\author{Eric G. Brown}
\email{e9brown@uwaterloo.ca}
\affiliation{Department of Physics and Astronomy, University of Waterloo, Waterloo, Ontario N2L 3G1, Canada}
\author{Robert B. Mann}
\email{rbmann@uwaterloo.ca}
\affiliation{Department of Physics and Astronomy, University of Waterloo, Waterloo, Ontario N2L 3G1, Canada}
\author{Eduardo Mart\'{i}n-Mart\'{i}nez}
\email{emartinm@uwaterloo.ca}
\affiliation{Institute for Quantum Computing, University of Waterloo, Waterloo, Ontario, N2L 3G1, Canada}
\affiliation{Department of Applied Mathematics, University of Waterloo, Waterloo, Ontario, N2L 3G1, Canada}
\affiliation{Perimeter Institute for Theoretical Physics, Waterloo, Ontario, N2L 2Y5, Canada}

\begin{abstract}

We explore the effects of different boundary conditions and coupling schemes on the response of a particle detector undergoing uniform acceleration in optical cavities. We analyze the thermalization properties of the accelerated detector via non-perturbative calculations. We prove non-perturbatively that if the switching process is smooth enough,  the detector thermalizes to the Unruh temperature regardless of the boundary conditions and the form of the coupling considered.  
\end{abstract}

\maketitle

\section{Introduction}

The interaction between matter and the gravitational field has evaded a complete quantum description since the first attempts to formulate a quantum theory of gravity more than 60 years ago. Lacking a satisfactory quantum description of the gravitational interaction, quantum field theory in curved spacetimes (which links general relativity with quantum field theory) is thus far the most satisfactory framework to describe the interaction of quantum fields with the space-time curvature.   As of today none of its predictions has been experimentally confirmed beyond analogue gravity  \cite{Barcelo2011},
and bringing those effects within experimental reach is a matter of great interest \cite{Chen1999a,Martin-Martinez2011,Rideout2012,AasenPRL}. 

One of the chief predictions of quantum field theory (QFT) in curved spacetimes is the well-known Unruh effect \cite{Unruh1976}. 
It dictates that a detector with constant acceleration $a$ in free space, in which the field is in the Minkowski vacuum, will experience a response equivalent to its submersion into a heat bath with a temperature proportional to its acceleration.
This phenomenon is intrinsically related to the so-called Hawking effect \cite{Hawking1974,Hawking1975}, 
and understanding it is essential in order to investigate more complex phenomena such as black hole dynamics and possible quantum corrections to relativistic gravity.

The first derivations of the Unruh effect (based on the characterization of the Minkowski vacuum in a unitarily inequivalent Rindler quantization scheme via Bogoliubov transformations) are not
above criticism.
A number of very strong assumptions have to be made in order to justify the observation of a thermal bath by an accelerated observer in the Minkowski vacuum.
For example, the
infinite amount of energy required to sustain the eternal Rindler trajectory.  It has also been argued that
difficulties arise in defining a Minkowski vacuum when boundary conditions are
specified on the scalar field on a manifold \cite{Narozhny2001}.

 Despite these criticisms,   later results in the context of axiomatic quantum field theory  \cite{Sewell1982} were used provide a model-independent derivation of the Unruh effect.  Furthermore, derivations based on accelerated particle detectors \cite{DeWitt1979} that appeared soon after the original derivation reassert the importance of the Unruh effect.  
The standard approach in grappling with the difficult
computational problem of an arbitrary system interacting with a scalar	
field on a curved spacetime manifold is to use an Unruh-DeWitt detector
\cite{DeWitt1979,Birrell1984}. This simplified detector model considers a two-level system coupled to a scalar field with a monopole interaction of the form 
\begin{equation}\label{eq:intH1}
H_I = \lambda(\tau)\,  \hat m\, \hat \phi[x(\tau)],
\end{equation}
where $\hat m$ is the monopole-moment of the detector, $\phi[x(\tau),t(\tau)]$ is the field operator evaluated along the worldline of the detector, and $\lambda(\tau)$ is a number-valued function that represents the strength and time-dependence of the coupling.
It has been shown that, although simple, this Hamiltonian is a good model of the light-matter interaction when no exchange of angular momentum is involved \cite{Martin-Martinez2012}.

Typically the Unruh-DeWitt model is used within the framework of perturbation theory, and very often restricted to lowest-order calculations (see, for example, \cite{Birrell1984,Louko2008,Barbado2012a}).
However, perturbation theory is not always applicable and breaks down when analyzing scenarios involving strong coupling, large energy, or large time scales.

While it is relatively easy to perturbatively show that the response of an accelerated detector to the vacuum state is Planckian \cite{Birrell1984},  perturbation theory is not the most appropriate approach to study the thermalization properties of the detector. 
In practice this is mainly because higher orders of
perturbation theory would be required, increasing the calculational complexity even beyond those of non-perturbative methods. More importantly, thermalization is an equilibrium result achieved over the course of long time scales.
In general, such time scales will not be accessible to perturbation theory since the perturbative parameter, i.e. $| \langle H \rangle \Delta T|$ becomes larger as time increases. Since thermalization is, in general, an equilibrium process that requires analysis in the limit $\Delta T\rightarrow \infty$, reasonable criticism may be raised about a perturbative claim of thermalization. 
Indeed, to check whether or not the detector evolves to an exactly thermal state, we will consider long time-scale evolution combined with adiabatic switching (to perturb the system the least when the interaction is switched on). 

Concretely, one should not only check that the probability of excitation of the detector has a Planckian response, one should also check to what extent the state of the detector becomes thermal if the detector is carefully switched on and if the interaction lasts for long enough times. 
This requires a complete calculation of the detector's density matrix; it is a common misconception that a detector's Planckian response implies that the detector thermalizes. 
For instance, the detector could evolve to a squeezed thermal state which may exhibit the same probability of excitation as some thermal state, but which is not actually a thermal state. 
By the use of non-perturbative methods we can make sure that thermalization is
achieved and that it is not an artifact of the use of perturbation theory in
regimes beyond its applicability.

Such non-perturbative methods were recently developed and applied to examine the response
of a detector within a cavity containing a scalar field \cite{Brown2012}.  
The cavity was a wave-guide with periodic boundary conditions in which the detector was allowed to entirely cycle several times during its evolution. Whereas this is physically reasonable for the case of periodic boundary conditions,  it is not the correct setting to compare with  more general boundary conditions.

We consider in this paper the thermality of accelerated detectors in optical cavities with different boundary conditions.
Extending  the work of Brown et al. \cite{Brown2012} (see also \cite{Bruschi:2012rx}), we will demonstrate non-perturbatively that an accelerated Unruh-DeWitt detector coupled to the vacuum state of a scalar field thermalizes to a temperature proportional to its acceleration, regardless of the boundary conditions imposed. The scenarios we consider here also differ from  previous work in the way that the detector trajectories are defined with respect to the cavity.  For example, we modify the cavity length such that the detector remains inside a single cavity during its interaction with the field,  which is of capital importance for physicality in the case of non-periodic cavities.

We do note that there has been an effort to understand how imposing different boundary conditions
 modifies the response of detectors in non-inertial scenarios in free space. 
For example, work has been done in a very different context to examine the continuum Rindler
case \cite{Rovelli2012}, and a number of boundary conditions in Hartle-Hawking vacua have been studied \cite{Hodgkinson2012}. 
However, these studies do not tell us if the boundary
effects of a cavity will prevent a uniformly accelerating particle detector from thermalizing due to the Unruh effect. 
To our knowledge the only work that addresses this issue is the aforementioned paper by Brown et al. \cite{Brown2012} in the periodic cavity case.

In addition to the study of the universality of the Unruh effect and the existence of thermalization for different sets of boundary conditions,
we shall also briefly comment on the effect of different methods of coupling the detector
to the field. 
Typically, we couple the detector locally to the field through the
detector's monopole moment, $\hat{\mu}_M = \left( \hat{a}_d + \hat{a}_d^{\dagger}\right)$.
Here, we will also explore the effects of a different form of coupling,
namely the coupling of the detector's monopole moment to the momentum of the
field. 

Our findings indicate that in all of the scenarios under consideration, the
Unruh effect occurs. We observe that the detector achieves thermalization with temperature proportional to acceleration. 
Thus, not only does the Unruh effect occur inside a cavity (which imposes an
IR-cutoff on the field   and, furthermore, isolates the field in the cavity from the rest of the spacetime), it appears to occur independently of the details of this IR cutoff and of the spatial distribution of the cavity modes. 
This demonstrates that the Unruh effect, which many have argued relies on idealized details and thus cannot lead to thermalization \cite{Nezhadhaghighi2013}, is in fact a very general and universal phenomenon and that thermalization of particle detectors can be computed non-perturbatively.
Not only is this a remarkable result from a fundamental point of view,  it also gives hope to the possibility of an experimental realization of the Unruh effect in quantum optical settings,  where it has been shown that general relativistic scenarios like the one we study here can already be simulated \cite{Diegger}.


Our paper is organized as follows. In Sect. \ref{SettingSect} we discuss the physical setup of our system, elaborating on the differences between the various scenarios and boundary conditions considered. In Sect. \ref{GaussianSect} we explain the oscillator-detector model that we will be using in our study, as presented in \cite{Brown2012}, and go on to discuss how we solved for the evolution of the detector-field system. In Sect. \ref{MainSect} we present our results on the thermalization of the accelerating detector along with the linear dependence of its temperature on acceleration. Furthermore, we demonstrate that these results are largely independent of the boundary conditions imposed on the field. In Sect. \ref{ConSect} we finish with some concluding remarks.

\section{The setting}   \label{SettingSect}

\begin{figure}[htb!]
	\includegraphics[scale=0.7]{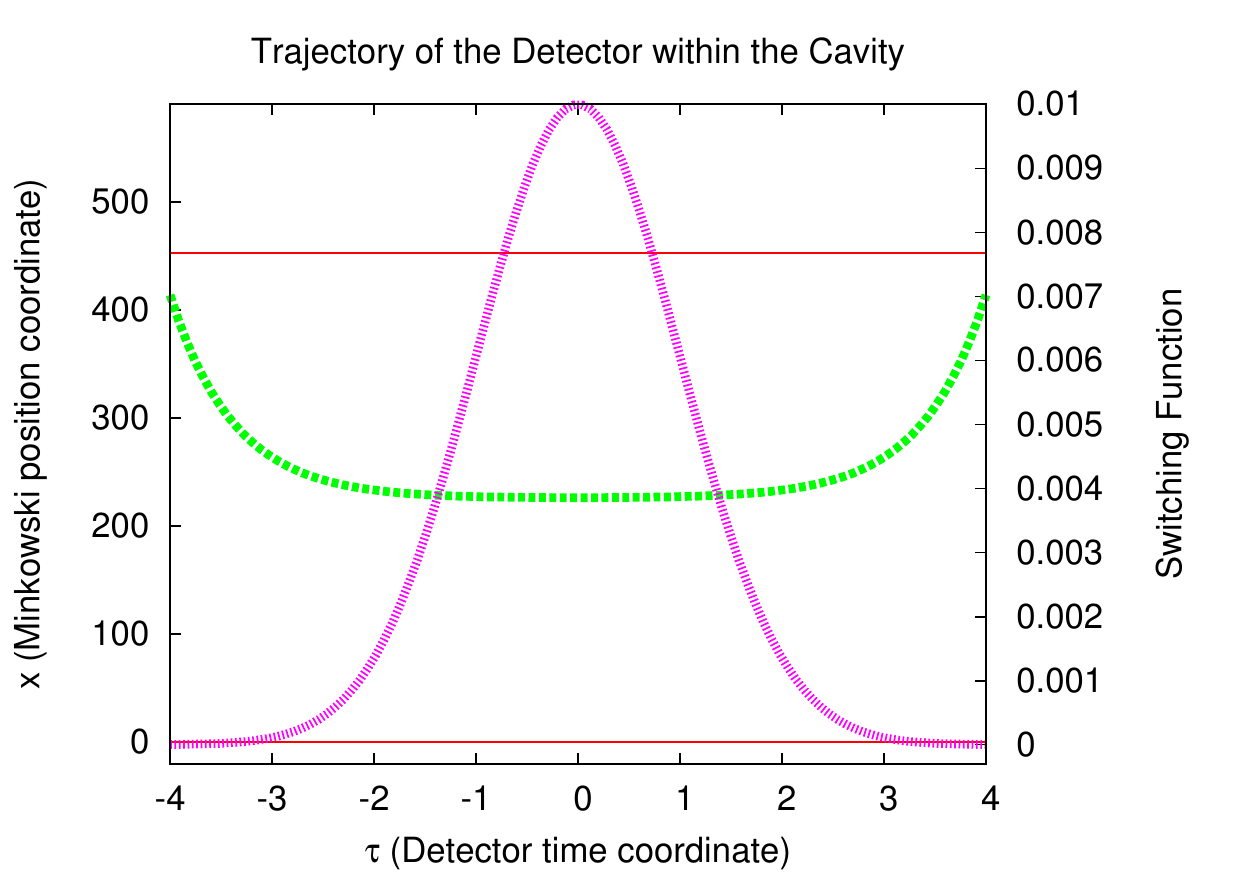}
	\caption{The detector's (green, dashed) trajectory through the cavity (red, solid) with acceleration $a=1.6$. The Gaussian switching function is plotted on the right axis with a jagged magenta line.}
	\label{figtraj}
\end{figure}

We will consider a uniformly accelerated point-like detector in its ground state going through a cavity prepared in the vacuum state.
The trajectory will be such that the detector starts moving inside the cavity with a given initial speed, with a constant acceleration in a direction opposite to its initial motion. Hence the atom will be decelerated while crossing the cavity.
The detector reaches the center of the cavity exactly when it reaches zero speed, and then travels back to the initial point with increasing speed until it reaches the position in which it started, having the same speed as when it entered the cavity but in the opposite direction. This trajectory of the atom within the cavity as a function of the detector's proper time $\tau$ is shown in Fig. \ref{figtraj}.
For larger values of the acceleration the detector will exit the cavity. In principle this is an issue as we mirror the field modes outside of the cavity.
However, over the range of accelerations that we will consider the coupling decays so quickly past the edges that these tails will not contribute significantly to the observed final state. In addition, we find that
the linearity of the temperature plots is preserved even when the detector escapes from the cavity.

To have a clean signature one must be careful with the way in which the detector is switched on \cite{Satz2007,Brown2012} since a sudden switching stimulates strong quantum fluctuations that may overcome the Unruh effect.
In order to reduce switching noise we apply the same approach as \cite{Brown2012}: 
the interaction is smoothly switched on following a Gaussian time profile so that switching quantum noise effects are reduced.
In particular, the switching function that we use has the form
\begin{equation}
	\lambda(\tau) = \lambda_0 \exp(-\tau^2/2 \delta^2).
	\label{eqn:gaussianswitch}
\end{equation}

We prepare the ground state of the detector and the vacuum of the field at a time $\tau=-T$ where the interaction is switched on following the Gaussian profile above, the atom has some initial speed and starts decelerating until it reaches the centre of the cavity at time $\tau=0$.
The atom continues accelerating until the time $\tau=T$ when it reaches the initial point again.
In the settings that we shall analyze, we will consider as parameters $T=4$, 
$\delta = 8/7$ and $\lambda_0 = 0.01$. 
With these parameters we find that the switching is more than smooth enough for our purpose; namely the detector's response from the switching noise is negligible compared with the response from acceleration.

We are going to consider different scenarios that correspond to different cavity field settings.
Let us rewrite the interaction Hamiltonian \eqref{eq:intH1} in the interaction picture in the following general form   \begin{align}\label{hamilgood}
\hat H_I&=\lambda(\tau)(\hat a_d e^{-\ii\Omega\tau}+\hat a^\dagger_d e^{\ii\Omega\tau}) \nonumber \\
&\times   \sum_n\!\Big(\hat a_n u_n[x(\tau),t(\tau)] +\hat a^\dagger_n u^*_n[x(\tau),t(\tau)] \Big)\,,
\end{align}
where we have expanded the field operator in terms of an orthonormal set of
field mode functions $u_n[x,t]$ which will depend on the boundary conditions
that are imposed upon the field. The normalization of the field modes is performed with respect to the Klein-Gordon inner product \cite{Birrell1984}.

If we choose a cavity of length $L$ and we impose the Dirichlet
boundary conditions $\phi[L,t]=\phi[0,t]=0$ (the two walls of the cavity are
ideal mirrors) we find that the mode functions are the stationary waves
\[u_n[x,t]=\frac{1}{\sqrt{k_n L}}e^{-\ii \omega_n t}\sin(k_n x)\]
where $n\in \mathbb{Z}^+$ and $\omega_n=k_n c=n\pi c/L$. 

In the case of a periodic cavity of length $L$ (which would correspond to physical settings such as closed optical fibres or microwave guides or any other setting with a torus topology), the modes are the set of right and left-moving waves
\[u_n[x,t]=\frac{1}{\sqrt{2 |k_n| L}}e^{-\ii (\omega_n t-k_n x)}\]
where $n$  is an integer, $k_n=2n\pi /L$ (negative (positive) $n$ corresponds to left-moving (right-moving) modes)  and $\omega_n=|k_n| c$. 

In the case of a Neumann cavity of length $L$ (this is to say, $\partial_x\phi[L,t]=\partial_x\phi[0,t]=0$), the modes become
\[u_n[x,t]=\frac{1}{\sqrt{k_n L}}e^{-\ii \omega_n t}\cos(k_n x)\]
where $n\in \mathbb{Z}^+$ and $\omega_n=k_n c=n\pi c/L$.

Finally, as we described above, the worldline of the detector inside the cavity parametrized in terms of its proper time will be given by 
\[t(\tau)=\frac{c}{a}\sinh(a\tau),\quad x(\tau)=\frac{L}{2}+\frac{c^2}{a}\left[\cosh(a\tau)-1\right]\]
and the interaction will be smoothly switched on following the curve
\eqref{eqn:gaussianswitch} from time $-T$ to $T$ with suitable values of $L$ and
$T$ such that the atom always remains in the cavity while the interaction is
``on'' (see Fig. \ref{figtraj}).

\section{Gaussian Formalism}  \label{GaussianSect}

We will introduce the non-perturbative oscillator-detector model. 
Replacing the usual two-level system in the Unruh-DeWitt model with a
harmonic oscillator is somewhat common in the literature
\cite{Unruh1989,Hu1994,Massar2006,Lin2007,Martin-Martinez2011,Dragan2011}.
However, in almost all of these cases, a perturbative approach was used, and not
many practical non-perturbative results have been obtained in the past. 
Here we will use the powerful non-perturbative Gaussian formalism developed in \cite{Brown2012,Bruschi:2012rx} to analyze the thermalization properties of the detector.

The only restrictions that apply to this formalism in a cavity scenario are that
the initial state of the system is a Gaussian state (such as the vacuum, a
coherent or squeezed state or a thermal state) and that the interaction
Hamiltonian is quadratic in the quadrature operators (in order to preserve the state's Gaussianity through time evolution).
Restricting our consideration to quadratic Hamiltonians is quite
reasonable since the interaction between matter and light is of this nature \cite{Scully1997}.
  
In addition, as the Unruh effect has almost solely been studied in the context of free space, this provides an excellent excuse for us to tread into unknown waters by considering different cavity settings.
Furthermore, any experimental verification of the Unruh effect is likely to
be  more easily implementable  in the context of optical cavities, so it is important to understand the phenomenon in such a scenario.

Let us summarize the tools that we are going to employ in this paper to obtain, non-perturbatively, the response of a particle detector to the field vacuum.
Instead of the interaction picture, it will be more convenient to work in the Heisenberg picture, following \cite{Brown2012}.
Let us form a vector from the detector's and field's annihilation and creation operators in the Heisenberg picture:
\begin{equation}
 	\hat{\bm{a}} \equiv ( \hat{a}_{d}, \hat{a}_{d}^{\dagger}, \hat{a}_1, \hat{a}_1^{\dagger}, \hat{a}_2,%
	\hat{a}_2^{\dagger}, \ldots, \hat{a}_N, \hat{a}_N^{\dagger})^T,
	\label{eqn:gqivec}
\end{equation}
where the subscript $d$ corresponds to the detector and the others correspond to the modes of the field.
The commutators of the components of this vector generate a symplectic form
\begin{equation}
	\bm{\Omega} \equiv \left( \begin{array}{c c c c c}
		0 & 1 & 0 & \hdots & 0 \\
		-1 & 0 & 1 & \hdots & 0 \\
		\vdots & \vdots & \vdots & \ddots & \vdots \\
		0 & \hdots & 0 & -1 & 0
	\end{array} \right) = \left[ \hat{\bm{a}}_i, \hat{\bm{a}}_j \right]
	\label{eqn:sympform}
\end{equation}
Similarly, we can form a vector of quadrature operators of the form
\begin{align}
	\hat{\bm{x}}=(\hat{q}_d,\hat{p}_d, \hat{q}_1, \hat{p}_1, \dots, \hat{q}_N,\hat{p}_N)^T.
	\label{eqn:phaseform}
\end{align}
These operators are related to the creation and annihilation operators of each mode by
\begin{align}
	\hat{q}_i=\frac{1}{\sqrt{2}}(\hat{a}_i+\hat{a}^\dagger_i), \;\;\;\; \hat{p}_i=\frac{i}{\sqrt{2}}(\hat{a}_i^\dagger-\hat{a}_i).
\end{align}

In order to ensure that Gaussian states will remain Gaussian over the course of their evolution,
we also need to evolve by a quadratic Hamiltonian:
\begin{equation}
 \hat{H} =  \Omega_d \hat{a}_d^{\dagger} \hat{a}_d +\frac{d t}{d \tau} \sum_n  \omega_n \hat{a}_n^{\dagger} \hat{a}_n + \hat{H}_I(\tau)
	\label{eqn:genham}
\end{equation}
where $\hat{H}_I$ will in our case be given  by (\ref{eq:intH1}) in the Heisenberg picture.
Here $\Omega_d$ is the frequency of the oscillator-detector and $\omega_n$ is the frequency of the $n^\text{th}$ field mode.
In our notation we use $\tau$ to denote the proper time of the detector (which is generally moving with respect to the cavity) and $t$ to denote the lab time, which is the time with respect to which the field evolves. 
When computing the system's evolution as generated by some Hamiltonian we must be careful to choose a specific time parameter and construct the Hamiltonian accordingly.
In the above Hamiltonian we choose to evolve with respect to the detector's proper time $\tau$. 
Since the field evolves with respect to $t$, this means that we must include a ``blue-shift factor'' on the free field's Hamiltonian. 
A more complete and rigorous explanation of this can be found in \cite{Brown2012}.

Since it is Gaussian, the state of the detector-field system can be completely described by a covariance matrix consisting of the first and second moments of the quadrature operators. In our scenario we need not consider states with first moments other than zero because of the absence of Hamiltonian terms linear in the quadrature operators, and so we determine the state by a covariance matrix of the form
\begin{equation}
	\sigma_{i j}=\langle \hat{x}_i \hat{x}_j+\hat{x}_j \hat{x}_i \rangle.
	\label{eqn:covmat}
\end{equation}
where $\hat{\bm{x}}$ is the vector formed by the dimensionless position and momentum
operators from equation \eqref{eqn:phaseform}. Thus the state of our single detector can be completely described using the $2 \times 2$ covariance matrix of the form 
\begin{equation}
	\bm{\sigma}_{d} \equiv\left( \begin{array}{c c}
		\langle \hat{q}_d^2 \rangle & \langle \hat{q}_d\hat{p}_d + \hat{p}_d\hat{q}_d \rangle \\
		\langle \hat{q}_d\hat{p}_d + \hat{p}_d\hat{q}_d \rangle & \langle \hat{p}_d^2 \rangle \\
	\end{array} \right)
	\label{eqn:covmat2}
\end{equation}

The time evolution of the entire covariance matrix, including both the detector and the field, is
governed by the equation of unitary evolution \cite{Brown2012}
\begin{equation}
	\bm{\sigma}(\tau) = \bm{S}(\tau) \bm{\sigma}_0 \bm{S}(\tau)^T
	\label{eqn:sigmaevol}
\end{equation}
where $\bm{S}$ is a symplectic matrix: $\bm{S}\bm{\Omega}\bm{S}^T=\bm{S}^T\bm{\Omega}\bm{S}=\bm{\Omega}$.

In addition, the symplectic matrix generated by a (generally time-dependent) Hamiltonian $\hat{H}(\tau)$ satisfies the equation
\begin{equation}
	\frac{d}{d \tau}\bm{S}(\tau) = \bm{\Omega} \bm{F}^{\text{sym}}(\tau)  \bm{S}(\tau)
	\label{eqn:timeevol}
\end{equation}
with initial condition $\bm{S}(0)=\bm{I}$. Here $\bm{F}^\text{sym}=\bm{F}+\bm{F}^T$, where $\bm{F}$ is a phase-space matrix encoding the form of the Hamiltonian via
\begin{align}
	\hat{H}(\tau)=\hat{\bm{x}}^T \bm{F}(\tau) \hat{\bm{x}}.
\end{align}

Keeping in mind that we will continue working in the Heisenberg picture (our
operators are fully time dependent), we will make use of some computational
techniques beyond what was indicated in \cite{Brown2012} that are inspired by
the principle of the interaction picture.  
Here, we make use of an exact solution of the free symplectic time-evolution matrix to speed up the computation. 

Namely, we split the evolution matrix into an exactly solvable part (we could call it the free part) and a non-exact part (that we could call the interaction part).
From equation (\ref{eqn:timeevol}), we take
\begin{equation}
	\bm{\Omega} \bm{F}^\text{sym}(\tau) \equiv \bm{K}_0 + \bm{K}_1(\tau)
	\label{eqn:split}
\end{equation}
where $\bm{K}_0$ is exactly solvable. For our circumstances, we choose
\begin{equation}
	\bm{K}_0 \equiv \bm{\Omega} \left(\bm{ F}^\text{sym}_{d} + \frac{d t}{d \tau} \bm{F}^\text{sym}_{f} \right).
	\label{eqn:k0}
\end{equation}
Here $\bm{ F}^\text{sym}_{d}$ and $\bm{F}^\text{sym}_{f}$ are the symmetrized matrices corresponding to the free Hamiltonians of the detector and field, respectively. That is, their non-symmetrized versions satisfy $\Omega_d \hat{a}_d^\dagger \hat{a}_d= \hat{\bm{x}}^T \bm{F}_d \hat{\bm{x}}$ and $ \sum_n \omega_n \hat{a}_n^{\dagger} \hat{a}_n =\hat{\bm{x}}^T \bm{F}_f \hat{\bm{x}}$.

Although, strictly speaking, $\bm{K}_0$ is time-dependent due to the $d t / d \tau$ factor, the fact that this is a total derivative and that we are integrating over $\tau$ to solve the dynamics means that we are still able to solve for the free evolution exactly,
so that if $\bm{K}_1(t) = 0$, equation (\ref{eqn:timeevol}) has the exact solution
\begin{equation}
	\bm{S}_0(\tau) = \exp \left[\bm{\Omega} \left(\bm{ F}^\text{sym}_{d} \tau + \bm{F}^\text{sym}_{f} t(\tau) \right)\right]
	\label{eqn:s0exact}
\end{equation}

Applying this interaction-picture-like approach, we define
\begin{align}
	\bm{S}^I(\tau) &\equiv \bm{S}_0^{-1}(\tau) \bm{S}(\tau) \nonumber \\
	\bm{K}_1^I(\tau) &\equiv  \bm{S}_0^{-1}(\tau) \bm{K}_1(\tau) \bm{S}_0(\tau)
	\label{eqn:interactiondef}
\end{align}
It is then easily seen that (\ref{eqn:timeevol}) becomes
\begin{align*}
	\frac{d \bm{S}^I(\tau)}{d\tau} = \bm{K}_1^{I}(\tau) \bm{S}^I(\tau).
\end{align*}

The evaluation of $\bm{S}^I(\tau)$ can then be accomplished by standard numerical techniques. The full Heisenberg evolution matrix is then simply $\bm{S}(\tau)=\bm{S}_0(\tau) \bm{S}^I (\tau)$, 
and the evolved state of the detector-field system is given by $\bm{\sigma}(\tau)=\bm{S}(\tau) \bm{\sigma}_0 \bm{S}(\tau)^T$.

\section{Unruh Temperature and Thermalization}   \label{MainSect}

Here we present the results obtained by applying the formalism of Sect. \ref{GaussianSect} 
to the scenario outlined in Sect. \ref{SettingSect}. Our goal
is to test the universality of the Unruh effect with respect to a change of
boundary conditions in the cavity.
We begin by considering an oscillator detector uniformly accelerating through a
cavity field that is initially in the vacuum state. 

It should be noted that when considering accelerating detectors it is necessary to include many modes in the field expansion.
This is due to the fact that the detector will experience an exponentially changing time dilation with respect to the cavity frame which translates into a modulating blueshift of the mode frequencies as seen by the detector.
Thus, the field modes with which the detector is resonant will rapidly change and, if the acceleration or time of evolution is large enough, very high mode numbers can make significant contributions to the evolution of the detector.
In our work we have been vigilant to ensure that enough field modes were included such that further additions do not modify the results obtained for the detector.

Because the non-perturbative Gaussian formalism presented in Sect. \ref{GaussianSect} slows down considerably for a large number of 
field modes, we use the standard perturbative formalism up to the first order to ensure complete convergence with respect to the number of field modes.
For example, in the periodic case this method involves evaluating the integral
\begin{equation}
	\int_{-T}^{T} d\tau \hat{H}_I(\tau)
	\label{eqn:inthamil}
\end{equation}
so, for periodic boundary conditions we therefore need to evaluate
\begin{equation}
	I_{n,\epsilon} = \lambda_0 \int_{-T}^{T} dt \cdot e^{ i \left( \Omega t  - \frac{2 \pi n \epsilon}{L} \exp(-\epsilon a t) \right) - \frac{t^2}{2 \sigma^2} }
	\label{eqn:perturbative}
\end{equation}
which is the contribution to the excitation probability amplitude for a given field mode, where $\epsilon$ is used to sum over left- and right-moving modes and the last term in the exponential is the Gaussian switching function (see below for more details).
We can use them (after normalization) in a partial sum over the number of field modes. That is,
\begin{equation}
	P = \sum_{n,\epsilon} \frac{1}{4 n \pi} | I_{n,\epsilon} |^2
	\label{eqn:perturbative2}
\end{equation}
This gives the first-order perturbation theory result for the probability of transition of the detector from the ground state to the
first excited state.

Now, when beginning its evolution the detector is initially in its ground state, but through its interaction with the field it will generally become excited.
After the evolution of the detector-field system is complete we will examine the state of the detector, which will be fully specified by its $2\times 2$ covariance matrix $\bm{\sigma}_d$.
In order to conclude that the detector has experienced a thermal Unruh bath
during its acceleration, we look for two things: first, that the detector has evolved to a thermal state and, second, that the corresponding temperature grows linearly with the acceleration experienced by the detector.

\subsection{Thermality}

The non-perturbative approach from \cite{Brown2012} is very well suited for testing the thermality of our detector.
Not only is thermality easily tested, but because we have the exact
(non-perturbative) state of the detector we are able to make a definitive
statement about thermalization. 
In order to make conclusions about
thermalization using perturbation theory, one would at the least need to expand to higher orders, 
and at the worst it would be impossible due to the long time scales typically required for thermalization to occur  where perturbation theory may break down.

Once the detector has completed its evolution it will be, up to phase-space rotation (i.e. free evolution), in a squeezed thermal state of the form
\begin{align}
	\bm{\sigma}_d=
	\begin{pmatrix}
		\nu e^r & 0 \\
		0 & \nu e^{-r}
	\end{pmatrix},
\end{align}
where $\nu$ is the covariance matrix's symplectic eigenvalue and $r$ is its squeezing parameter.
These diagonal entries are just the eigenvalues of $\bm{\sigma}_d$, $\lambda_{\pm}=\nu e^{\pm r}$, from which the symplectic eigenvalue and squeezing parameter follow as $\nu=\sqrt{\lambda_{+} \lambda_{-}}$ and $e^{2r}=\lambda_+ / \lambda_-$.
We must now determine whether the amount of thermality introduced by $\nu$ is a much greater contributor to the energy of the detector's state compared to the amount of squeezing. 
If so, then the state can be said to be nearly thermal. 
If they are comparable, or if squeezing has the greater contribution, then we can not claim that the detector thermalizes.

To compare these two effects we study how they contribute to the free energy of the detector.
For small squeezing (which is satisfied in our scenario) the energy above the
ground state energy, to leading order in $r$,  a  power series expansion straightforwardly yields
\begin{align}
	E-E_0=\Omega_d \left[(\nu-1)+\frac{1}{2}\nu r^2\right],
\end{align}
where $\Omega_d$ is the detector frequency.
Since $\nu$ is of order unity (and is in fact remains very close to unity for our situation) a good test for thermality is 
\begin{equation}
	\nu-1 \gg r^2
	\label{eqn:thermaltest}
\end{equation} 
If this inequality is satisfied then the detector can be said to be very nearly thermal.
Equivalently, if
\begin{equation*}
	\delta \equiv \frac{r^2}{\nu - 1}
\end{equation*}
is very small, the detector is said to be thermal.

We find that the detector thermalizes very well in all of the three boundary conditions considered: periodic, Dirichlet, and Neumann. We find numerically that for the parameters given in Sect. \ref{SettingSect}, $\delta$ is on the order of $10^{-6}$ in all three cases.
That is, the squeezing experienced by the oscillator is extremely minute
compared to its thermality, and thus the detector can be said to be very nearly thermal.

The first of our two conditions to verifying the Unruh
effect (thermality and temperature proportional to acceleration) is satisfied
for all three boundary conditions.

\subsection{Unruh temperature}

We are now in a position to compute the temperature of the evolved detector $\bm{\sigma}_d$. For a single oscillator of frequency $\Omega_d$ the form of an exactly thermal state is $\bm{\sigma}_d^\text{therm}=\text{diag}(\nu,\nu)$, for which the temperature is \cite{Adesso2007}
\begin{align}\label{tempnon}
	T=\Omega_d \left[\ln \left(1+ \frac{2}{\nu-1} \right) \right]^{-1}.
\end{align}
Since in our scenario we have already confirmed that our detector thermalizes to an excellent approximation we are able to use this equation to compute the temperature of our detector with negligible error, where $\nu=\sqrt{\lambda_+ \lambda_-}$ as above.

\begin{figure*}[htb!]
	\includegraphics[scale=0.91]{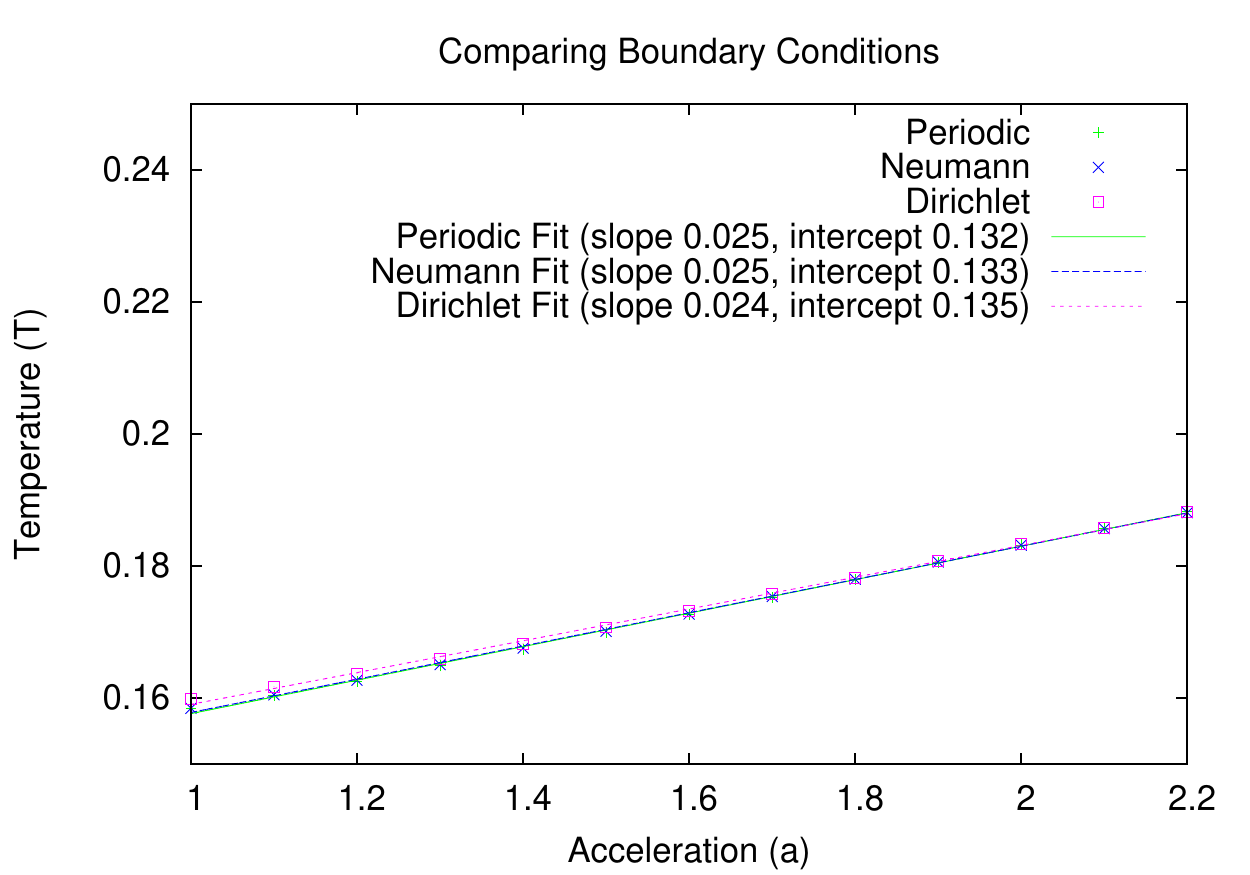}
	\caption{Comparison of different boundary conditions on the detector's temperature. The least-squares fitted results are perturbative with 9000 modes considered (evaluated as in equation (\ref{eqn:perturbative2})),
	while the non-perturbative data (not plotted) used 240 field modes and a tolerance of $10^{-11}$. The cavity length was chosen to be $L=144 \pi$.}
	\label{boundarycomp}
\end{figure*}

For each of the three boundary conditions (periodic, Dirichlet, and Neumann) we
compute this temperature for various values of acceleration. These results
are displayed in figure (\ref{boundarycomp}).
Notice that our least-squares fit is performed on first-order \emph{perturbative}
results. For small coupling strength the perturbative result for the transition probability of the accelerated detector
is in close agreement with the result obtained by the non-perturbative approach. Specifically, the probability computed up to leading order (as explained above) is of $\mathcal{O}(\lambda^2)$ and it is easily shown that the next relevant order, and thus the difference between the perturbative and non-perturbative answers, is  $\mathcal{O}(\lambda^4)$. For small temperatures, therefore, one need not use the non-perturbative approach to estimate the temperature of the oscillator detector, assuming that the detector is in a thermal state. When including a very large number of field modes it is computationally more convenient 
to plot the first order perturbative results with three objectives in mind: 1) from the probability (\ref{eqn:perturbative2}) and assuming thermality, compute the temperature using the standard Boltzmann distribution and check that it varies linearly with acceleration,  2) check that the non-perturbative 
results (using (\ref{tempnon})) are computed with enough numerical accuracy to reproduce the perturbative plot up to $\mathcal{O}(\lambda^4)$, ensuring that
had we used an infinite number of modes both methods would converge, and 3) use the non-perturbative results
to ensure thermality of the system. Note that this last step  cannot be easily carried out with a perturbative calculation (this was done and discussed in the previous section).

From the plot of the
trajectory (Fig. \ref{figtraj}), notice that very high acceleration data points involve the atoms exiting the cavity for a small part of their trajectories. To prevent
switching noise from damaging the results, we keep the field continuous beyond the cavity, but because
the switching function is effectively zero when the detector crosses the boundary, the interaction will
be negligible. Notice that the high acceleration results are not qualitatively different from lower acceleration results, justifying our 
negligibility assumption.
 
\begin{figure*}[htb!]
	\includegraphics[scale=0.91]{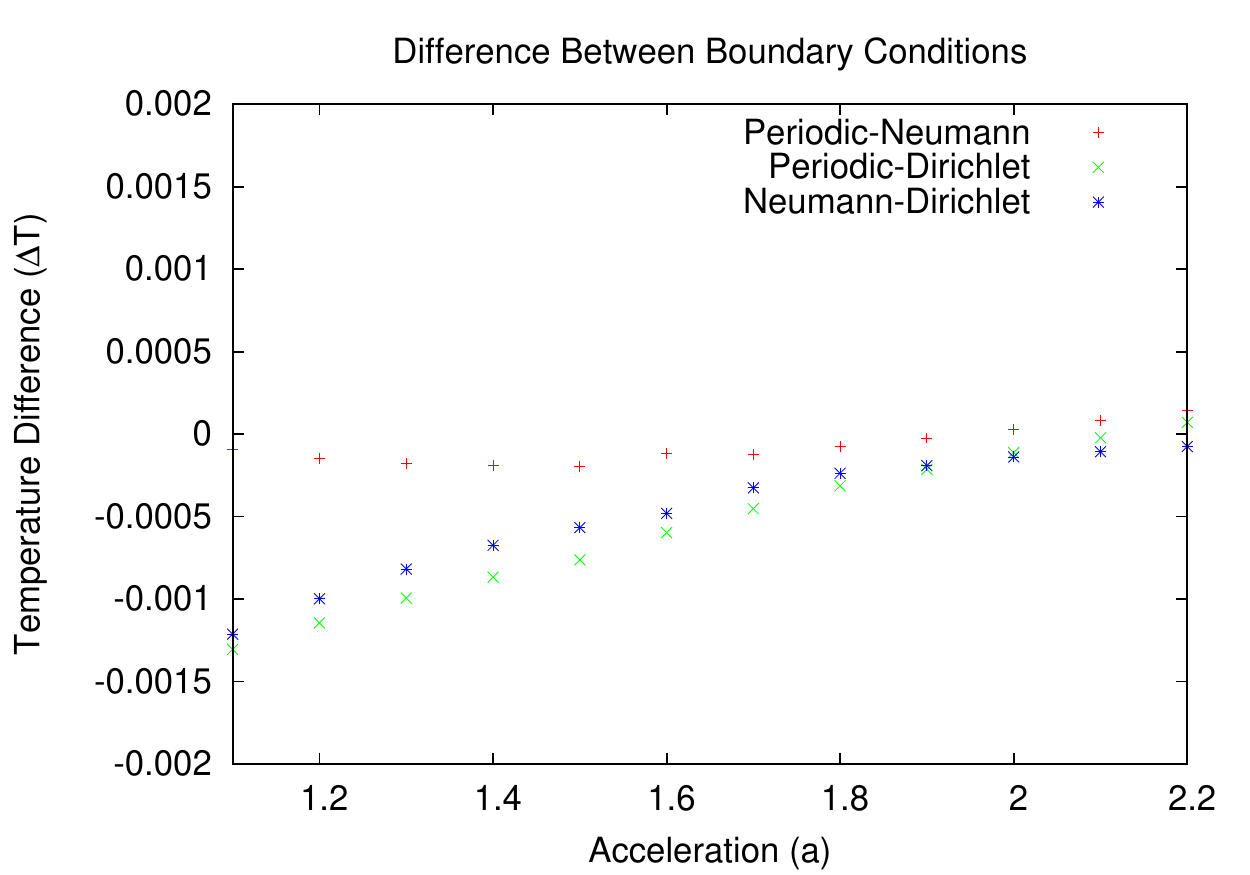}
	\caption{The difference between the perturbative curves in Figure \ref{boundarycomp}.}
	\label{boundarycomp2}
\end{figure*}

Remarkably we find that for all three boundary conditions the temperature grows linearly with acceleration, demonstrating that the qualitative features expected of the Unruh effect are very much independent of the details of the cavity. 
This settles any  doubt regarding the existence of the Unruh effect when an IR-cutoff for the field is introduced (i.e. when inside a cavity). There has been some skepticism \cite{Fedotov:2002hc,Narozhny:2004xx,Ford:2005sh} stemming from the large number of technical assumptions that go into the canonical derivation of the Unruh effect \cite{Unruh1976} and how the presence of a cavity might alter or even eliminate its existence. We have demonstrated not only thermality and the existence of the effect in a cavity
 (first shown in \cite{Brown2012} and reaffirmed here), but also  that the boundary conditions ascribed to this cavity are all but irrelevant (see Fig. \ref{boundarycomp2}). Indeed the numerical similarity between the different cases is striking.   

We note that the slope of the detector temperature with respect to acceleration is not equal to the value of $1/2\pi$ predicted by the canonical free-space derivation. This is not overly surprising since we are working in a cavity setting rather than free space; significant border effects should be expected when studying such phenomena. Our results demonstrate, however, that the inclusion of an IR cutoff does not destroy the Unruh effect understood as the thermal response of a particle detector with a temperature proportional to the acceleration, and what is more, that the detector actually thermalizes  to that particular temperature. If we were to take our cavity to the continuum limit we would expect (at least in the case of periodic boundary conditions) the slope to converge to the usual value of $1/2\pi$. We leave such a study for future work.

\subsection{XP Coupling}

In this portion of the paper we shall briefly discuss the effect of varying the form of
coupling between the field and the detector. In particular, the question we ask
is: what happens when the monopole coupling of equation (\ref{eq:intH1}) is
modified such that the detector couples to the conjugate momentum of the field instead of the field itself?

In this form, we can no longer maintain the pointlike coupling assumption for the detector-field coupling. After a Fourier transform, it is
apparent that a point-like $X-P$ coupling is akin to coupling the detector's monopole moment to the
field in an extremely delocalized way, such that the detector couples to the field
everywhere.  Thus this coupling is ill-defined and, without the introduction of  spatial smearing, 
will yield divergences.

We will therefore regularize the interaction assuming that the detector's coupling strength to the canonical momentum of the field varies with the field frequency. A frequency-dependent effective coupling appears naturally when considering spatially smeared detectors \cite{Schlicht2004,Martin-Martinez2012}, although in our case we would make the simpler assumption that the coupling strength is inversely proportional to the frequency of the mode.
In this way we can analyze the following coupling for periodic boundary conditions:
\begin{align}
	H_{int} = \ii\sum_{n} &  \lambda_n \cdot \sqrt{\frac{\omega_n}{2 L}} \left( a_d + a_d^{\dagger} \right) \nonumber \\
		& \left[ a_n e^{i k_n x(\tau)} - a_n^{\dagger} e^{-i k_n x(\tau)} \right]
	\label{eqn:xpcoupling}
\end{align}
where $\lambda_n \rightarrow   \lambda(\tau)/{\omega_n}$ so that
the energy density falls off with high energy modes at the same rate as in the 
$X-X$ coupling case. We note here that in first order perturbation theory,
the negative sign does not contribute to the probability of transition and this scenario is
exactly the same as the $X-P$ coupling.

In the non-perturbative case, we produced the same plot as the periodic curve of figure \ref{boundarycomp}
with the modified coupling; all data points were exactly the same. This tells us that the periodic $X-P$
coupling sign change is still a symmetry of the non-perturbative case. We believe that this should be observable
when examining the equations (\ref{hamilgood}) and (\ref{eqn:xpcoupling}) but we do not prove this here.

\section{Conclusions}   \label{ConSect}

In this work we have non-perturbatively solved for the evolution of an oscillator detector undergoing uniform acceleration through a cavity field. We have confirmed recent previous work  \cite{Brown2012} demonstrating 
 that the Unruh effect does indeed occur inside a cavity, and furthermore we have demonstrated that this result is independent of the boundary conditions applied to the field, implying that the Unruh effect is a very universal phenomenon.
Specifically, we have considered vacuum cavity fields with periodic, Dirichlet, and Neumann boundary conditions.
In all three cases we have observed that an accelerating oscillator detector evolves to a thermal state and that the temperature obtained by the detector increases linearly with its acceleration.
Furthermore the results between the three cases are numerically very similar.
This indicates that not only is the phenomenon qualitatively universal but, with respect to the case of boundary conditions, also quantitatively universal.
We have also made some conclusions regarding the use of different detector-field couplings that further strengthens these claims.

Moreover, our use of the non-perturbative oscillator model has allowed us to make significantly stronger claims regarding the thermality experienced by the detector than can be made using the standard perturbative framework typically employed in the literature.
That is, we have concluded that in our scenario an accelerating detector in fact evolves to a thermal state, rather than merely exhibiting a thermal response function.
Questions of thermalization cannot be made in perturbation theory without resorting to higher order expansions, and in some scenarios may actually be impossible due to the large time scales often required for thermalization where perturbation theory breaks down.

More generally, the results of this paper suggest that the Unruh effect and similar phenomena such as Hawking radiation may be largely independent of the details of the system \cite{Crispino2008}. In addition to theoretical interest, such universality bodes well for   an eventual experiment where the Unruh effect could be measured.
 
\section{Acknowledgments}
 The authors would like to thank very much William Donnelly for his helpful comments and insight on the interaction picture approach to the Gaussian formalism.  This work was supported in part by the National Sciences and Engineering
Research Council of Canada. E. M-M. was partially funded by the Banting Postdoctoral Fellowship Programme. W. B. was funded by the Vanier CGS Award.


\end{document}